\documentclass[9pt]{article}
\usepackage[preprint]{spconf}
\usepackage{array}
\usepackage{multirow}
\usepackage{color}
\usepackage{amsmath,amsfonts}
\usepackage{graphicx}
\usepackage{booktabs}
\usepackage{cite}

\copyrightnotice{978-1-5386-8151-0/18/\$31.00~\copyright2018 IEEE}

\title{SPEECH SEPARATION USING PARTIALLY ASYNCHRONOUS\\MICROPHONE ARRAYS WITHOUT RESAMPLING}

\name{Ryan M. Corey and Andrew C. Singer \thanks{This material is based upon work supported by the National Science Foundation Graduate Research Fellowship Program under Grant Number DGE-1144245.}}
\address{University of Illinois at Urbana-Champaign}

\begin{document}
\ninept

\maketitle
\begin{abstract}
We consider the problem of separating speech sources captured by multiple spatially
separated devices, each of which has multiple microphones and
samples its signals at a slightly different rate. Most asynchronous
array processing methods rely on sample rate offset estimation and
resampling, but these offsets can be difficult to estimate if the
sources or microphones are moving. We propose a source separation
method that does not require offset estimation or signal resampling. 
Instead, we divide the distributed array into several
synchronous subarrays. All arrays are used jointly to estimate the
time-varying signal statistics, and those statistics are used to design
separate time-varying spatial filters in each array. We demonstrate
the method for speech mixtures recorded on both stationary and moving
microphone arrays.
\end{abstract}

\begin{keywords}
Asynchronous microphone array, ad hoc microphone array, distributed arrays, sampling rate
offset, audio source separation, spatial filtering, speech enhancement
\end{keywords}

\section{Introduction}

Microphone arrays are useful for separating and enhancing audio signals
because they can isolate sound sources coming from different directions
\cite{gannot2017consolidated}. Over the last few years, microphones
have become ubiquitous in consumer electronic devices such as mobile
phones, hearing aids and other listening devices, computers, gaming
systems, and smart speakers. If many distributed microphones were
combined into a single \emph{ad hoc} array, they would provide greater
spatial resolution and therefore better separation performance than
any one of the devices alone \cite{doclo2009reduced,bertrand2011applications,taseska2014informed,miyabe2015blind,wang2016correlation,bahari2017blind,cherkassky2017blind,chiba2014amplitude,souden2014location}.

Microphones on different devices are sampled at slightly different
rates due to hardware variations. Although negligible in most applications,
these offsets can be critical in array processing, which relies on precise
phase relationships between microphones. Several asynchronous array processing methods have been proposed in the literature. In \cite{miyabe2015blind,wang2016correlation,bahari2017blind,cherkassky2017blind}, the systems first estimate the sample rate offsets and resample
the signals to a common rate. The resampled signals can then be combined
coherently using conventional array processing techniques. Unfortunately,
existing sample rate estimation algorithms are known to work poorly
for moving sources \cite{miyabe2015blind,wang2016correlation} and
often do not work at all for moving microphones, as we will demonstrate
in Section \ref{subsec:Sample-rate-offset}. In \cite{souden2014location,chiba2014amplitude}, the sources are separated using single-channel masks that do not require resampling, but also do not take full advantage of the spatial diversity afforded by arrays. To separate sources in the most challenging environments, we need new asynchronous source separation techniques that do not require resampling and that scale well to devices with many microphones.

\begin{figure}
\begin{centering}
\includegraphics{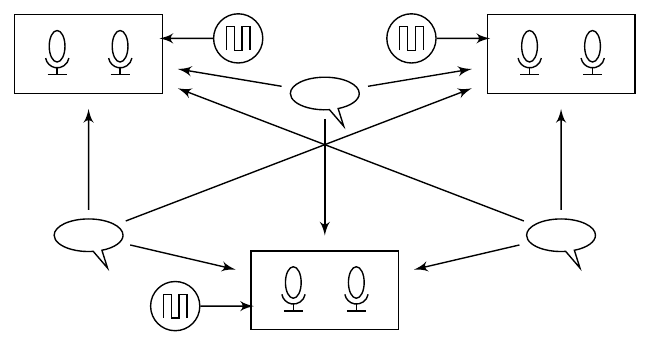}
\par\end{centering}
\caption{\label{fig:problem}We wish to separate $K$ sources using $M$ microphone arrays, each with its own sample clock.}
\end{figure}

In this contribution, we consider \emph{partially asynchronous} microphone
arrays in which some of the microphones do share a common sample
clock but others do not, as shown in Figure \ref{fig:problem}. As microphones have become smaller and less
expensive, many devices now include at least two. We can take advantage
of this partial synchronization to perform multimicrophone source
separation without resampling.
In our proposed system, each device applies a separate linear time-varying spatial filter \cite{corey2017underdetermined} to the signals collected by its local microphone array. The filter coefficients are computed using information about the source statistics from the full distributed array. For speech and other sparse signals, this shared information can take the form of source activity probabilities computed using spatial features from each array \cite{souden2014location}.
We demonstrate the proposed algorithm on real-world recordings of up to eight speech sources using both stationary and moving asynchronous microphone arrays.

\section{Asynchronous Array Processing}

Consider a set of $M$ distributed arrays and let $\mathbf{x}_{\mathrm{c},m}(t)$
be the vector of continuous-time signals captured by array $m$ for
$m=1,\dots,M$. The arrays need not have the same number of microphones. If the arrays shared a common sample period $T$,
then the sampled discrete-time sequences would be $\tilde{\mathbf{x}}_{\mathrm{d},m}[\tau]=\mathbf{x}_{\mathrm{c},m}(\tau T)$
for integer time indices $\tau$. Instead, each array $m$ has its own sample
period $T_{m}$, so that the sampled data vectors are $\mathbf{x}_{\mathrm{d},m}[\tau]=\mathbf{x}_{\mathrm{c},m}(\tau T_{m})$
for $m=1,\dots,M$. The received signals are due to $K$ independent
sound sources, so that 
\begin{equation}
\mathbf{x}_{\mathrm{d},m}[\tau]=\sum_{k=1}^{K}\mathbf{c}_{\mathrm{d},m,k}[\tau]\quad\text{for }m=1,\dots,M,
\end{equation}
where $\mathbf{c}_{\mathrm{d},m,k}[\tau]$ is the response of
array $m$ to source $k$, which is often called the source image \cite{vincent2006performance}.
The sources may include both directional sound sources and diffuse
noise. Our goal is to estimate one or more of the source images $\mathbf{c}_{\mathrm{d},m,k}[\tau]$
from the mixtures $\mathbf{x}_{\mathrm{d},1}[\tau],\dots,\mathbf{x}_{\mathrm{d},M}[\tau]$. 

\subsection{Sample rate offset model}
\label{subsec:Sample-rate-offset}

\begin{figure}
\begin{centering}
\includegraphics{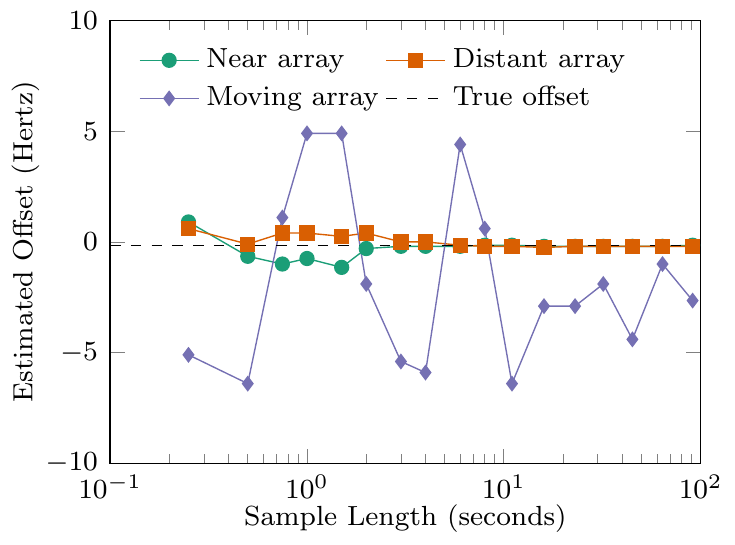}
\par\end{centering}
\caption{\label{fig:moving_sro}Estimated sample rate offsets between closely
spaced, distant, and moving arrays in an eight-talker cocktail
party scenario (see Section \ref{sec:Moving-Wearable-Arrays}) using handheld recorders and the two-stage correlation maximization algorithm
\cite{wang2016correlation}.}
\end{figure}

Let $\mathbf{c}_{m,k}[n,f]$, $\tilde{\mathbf{x}}_{m}[n,f]$,
and $\mathbf{x}_{m}[n,f]$ be the short-time Fourier transform
(STFT) vectors of the corresponding discrete-time sequences, where $n$ is the frame
index and $f$ is the frequency index. Since each array has a different
sample rate, the $[n,f]$ indices of each sequence $\mathbf{x}_{m}[n,f]$
correspond to slightly different continuous-time intervals and frequencies.
We assume that the sample times are coarsely synchronized and that the sample rate offsets are sufficiently small that the
sequences $\mathbf{x}_{\mathrm{d},m}[\tau]$ are offset from each
other by much less than one STFT frame length over the period
of interest. We can model the effect of those offsets by \cite{wang2016correlation}
\begin{equation}
\mathbf{x}_{m}[n,f]=e^{j\alpha_{m}[n,f]}\tilde{\mathbf{x}}_{m}[n,f],
\end{equation}
where $\alpha_{m}[n,f]$ is a phase shift due to the small sample
rate offset at array $m$. Then, assuming that the sequences are
zero-mean random processes, the across-array correlations are given by 
\begin{equation}
\mathbb{E}\left[\mathbf{x}_{m}[n,\!f]\mathbf{x}_{l}^{H}[n,\!f]\right]=\mathbb{E}\left[e^{j(\alpha_{m}[n,\!f]-\alpha_{l}[n,\!f])}\tilde{\mathbf{x}}_{m}[n,\!f]\tilde{\mathbf{x}}_{l}^{H}[n,\!f]\right],
\end{equation}
where $\mathbb{E}$ denotes expectation and $H$ the Hermitian transpose.
If the sample rate offsets are sufficiently small and time-invariant over the period of interest, then
each $\alpha_{m}[n,f]$ is approximately proportional to $nf(T^{-1}-T_{m}^{-1})$
\cite{wang2016correlation}. 

If the $\tilde{\mathbf{x}}_{m}[n,f]$
are approximately stationary over a long time interval, then the relative
sample rate offsets can be estimated based on these cross-correlations
\cite{miyabe2015blind,wang2016correlation,bahari2017blind} and the
$\mathbf{x}_{\mathrm{d},m}[\tau]$ sequences can be resampled
to obtain estimates of $\tilde{\mathbf{x}}_{\mathrm{d},m}[\tau]$. Correlation-based
methods are known to be sensitive to source motion \cite{miyabe2015blind,wang2016correlation}.
Movement of the microphones themselves is fatal, since sample rate
offsets and constant-velocity motion induce nearly identical linear
phase shifts \cite{cherkassky2014blind}. Figure \ref{fig:moving_sro}
shows the performance of a blind sample rate estimation algorithm
\cite{wang2016correlation} in a cocktail party scenario. It works
well when the microphones are stationary, even if they are far apart,
but it fails when one microphone moves relative to the other. Thus,
these algorithms are poorly suited to cocktail party scenarios with
microphones worn or carried by moving humans.

Here, we consider a worst-case scenario in which we know little about
the phase offsets between arrays. In particular, we model each $\alpha_{m}[n,f]$
as an independent random variable uniformly distributed from $-\pi$
to $\pi$. Under this model, since $\mathbb{E}\left[e^{j\alpha_{m}[n,f]}\right]=0$,
by linearity of expectation we have 
\begin{equation}
\mathbb{E}\left[\mathbf{x}_{m}[n,f]\mathbf{x}_{l}^{H}[n,f]\right]=0\text{\quad}\text{for }m\ne l.\label{eq:uncorrelated}
\end{equation}
The captured sequences are thus uncorrelated across arrays. Assuming
that the source images are uncorrelated with each other, their linear
minimum mean square error estimators are given by the multichannel
Wiener filters 
\begin{equation}
\hat{\mathbf{c}}_{m,k}[n,f]=\mathbf{R}_{m,k}[n,f]\left(\sum_{k=1}^{K}\mathbf{R}_{m,k}[n,f]\right)^{-1}\mathbf{x}_{m}[n,f],\label{eq:mwf}
\end{equation}
for $m=1,\dots,M$ and $k=1,\dots,K$, where each $\mathbf{R}_{m,k}[n,f]=\mathbb{E}\left[\mathbf{c}_{m,k}[n,f]\mathbf{c}_{m,k}^{H}[n,f]\right]$
is the time-varying source image covariance matrix. Since the images
are due to both directional and diffuse sources, we assume that
$\sum_{k=1}^{K}\mathbf{R}_{m,k}[n,f]$ is nonsingular for all
$m$, $n$, and $f$. Thus, the linear estimators for the source images at each array 
use only the local microphones in that array. If each array has only a
few microphones, then these filters might perform quite poorly compared
to those for a synchronous distributed array.

\subsection{Distributed spatial filtering}

\begin{figure}
\begin{centering}
\includegraphics{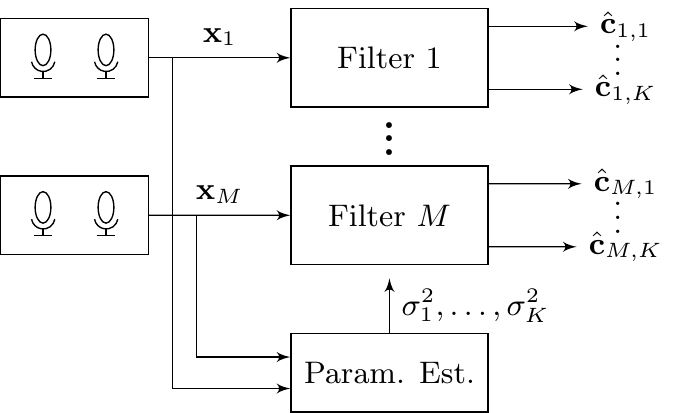}
\par\end{centering}
\caption{\label{fig:system}Each device estimates each source image using its
local microphones. The source powers are estimated using all $M$ arrays.}
\end{figure}

The multichannel Wiener filter (\ref{eq:mwf}) is often implemented
using time-varying estimates $\hat{\mathbf{R}}_{m,k}[n,f]$ of
the nonstationary source covariances \cite{taseska2014informed,souden2013multichannel,dmour2011gmm,duong2010fullrank,ozerov2010multichannel}.
Separation algorithms rely on good covariance estimates, and that is
where we can take advantage of the asynchronous arrays. Although the
sequences $\mathbf{x}_{m}[n,f]$ and $\mathbf{x}_{l}[n,f]$
are uncorrelated for $m\ne l$ due to their assumed-random phase shifts,
they are not independent: both are generated by the same set of sources.
Thus, we can use information from all $M$ arrays to estimate the
time-varying source statistics, then use those statistics to create
$M$ time-varying spatial filters. The proposed system is shown in
Figure \ref{fig:system}.

We will apply a variant of the full-rank local Gaussian model \cite{duong2010fullrank},
in which we assume that each source image $\mathbf{c}_{m,k}[n,f]$
has zero mean and a conditional normal distribution given its covariance
\begin{equation}
\mathbf{R}_{m,k}[n,f]=\sigma_{k}^{2}[n,f]\bar{\mathbf{R}}_{m,k}[f],
\end{equation}
where $\sigma_{k}^{2}[n,f]$ is the time-varying source spectrum and
$\bar{\mathbf{R}}_{m,k}[f]$ is the frequency-dependent spatial
covariance, which depends on the source and array geometry and room
acoustics. For simplicity, we assume here that each $\bar{\mathbf{R}}_{m,k}[f]$
is time-invariant and that the full-rank covariance matrix 
accounts for uncertainty due to motion of the array. As is typically done
with the local Gaussian model, we assume that the \textbf{$\mathbf{c}_{m,k}[n,f]$}
are conditionally independent across $n$, $f$, and $k$ given the
source spectra $\sigma_{1}^{2}[n,f],\dots,\sigma_{K}^{2}[n,f]$. Here,
we further assume conditional independence across $m$, which reflects
the uncorrelatedness of the array signals from (\ref{eq:uncorrelated}).

The proposed estimation method is as follows:
\begin{enumerate}
\item Estimate the spatial parameters $\bar{\mathbf{R}}_{m,k}[f]$ using
any suitable method. We show experimental results in Section \ref{sec:experiments}
using both a blind method and a method based on training.
\item Find estimates $\hat{\sigma}_{k}^{2}[n,f]$ of the time-varying source
spectra $\sigma_{k}^{2}[n,f]$ using the observations from all $M$
arrays. We propose an estimator for sparse mixtures in Section \ref{subsec:Joint-Spectral-Estimation}.
\item Use the estimated source spectra and spatial parameters in (\ref{eq:mwf})
to estimate the source images at each array:
\end{enumerate}
\begin{equation}
\hat{\mathbf{c}}_{m,k}[n,\!f]=\hat{\sigma}_{k}^{2}[n,\!f]\bar{\mathbf{R}}_{m,k}[f]\!\left(\sum_{s=1}^{K}\!\hat{\sigma}^2_{s}[n,\!f]\bar{\mathbf{R}}_{m,s}[f]\right)^{-1}\!\!\!\mathbf{x}_{m}[n,\!f]. \label{eq:est_mwf}
\end{equation}

\subsection{\label{subsec:Joint-Spectral-Estimation}Joint spectral estimation
for sparse sources}

There are many methods to estimate time-varying source spectra, such
as those based on expectation maximization \cite{dmour2011gmm,duong2010fullrank}
and nonnegative matrix factorization \cite{ozerov2010multichannel}.
Since here we are interested in speech sources, we will demonstrate a
classification method that takes advantage of the time-frequency sparsity
of speech \cite{rickard2002approximate}. The W-disjoint orthogonal
model, which is most often used for single-channel methods such as
time-frequency masks \cite{yilmaz2004duet} but has also been applied
for underdetermined multimicrophone separation \cite{taseska2014informed,araki2007multiple},
assumes that for every $[n,f]$, we can assign a state $s[n,f]\in\left\{ 1,\dots,K\right\} $
such that $\sigma_{s[n,f]}^{2}[n,f]\gg\sigma_{k}^{2}[n,f]$ for $s[n,f]\ne k$.
To account for periods with no active directional sources, we include at least
one stationary diffuse noise source in the model. 

Let $\sigma_{k|s}^{2}[f]$
denote the variance of source $k$ at frequency index $f$ when the system is in state $s$.
We model the variance as taking one of two values for each source, depending
on the state:
\begin{equation}
\sigma_{k|s}^{2}[f]=\begin{cases}
\sigma_{k,\text{high}}^{2}[f], & \text{if }k=s\\
\sigma_{k,\text{low}}^{2}[f], & \text{if }k\ne s.
\end{cases}
\end{equation}
Typical mask-based systems choose $\sigma_{k,\text{low}}^{2}=0$,
but since microphone arrays can steer multiple nulls at once,
it is advantageous to include all sources in the model. Here, we choose
$\sigma_{k,\text{high}}^{2}[f]$ and $\sigma_{k,\text{low}}^{2}[f]$
to be respectively 10 dB above and 10 dB below the long-term average
source spectrum, which we have found to work well for speech sources
\cite{corey2017underdetermined}. The diffuse noise source has the
same assumed spectrum in every state, and its magnitude can be tuned
to improve the conditioning of the matrices in (\ref{eq:est_mwf}). In our experiments in Section \ref{sec:experiments}, we use a spatially uncorrelated spectrum similar in power to that of the directional speech sources.

Under the local Gaussian model, the log-likelihood of the observations
in state $s$ is given by
\begin{align}
\log p_{s}[n,\!f]= & - \sum_{m=1}^{M}\mathbf{x}_{m}^{H}[n,\!f]\left(\sum_{k=1}^{K}\sigma_{k|s}^{2}[f]\bar{\mathbf{R}}_{m,k}[f]\right)^{-1}\mathbf{x}_{m}[n,\!f] \nonumber \\
& -\sum_{m=1}^{M} \log\det\left(\pi\sum_{k=1}^{K}\sigma_{k|s}^{2}(f)\bar{\mathbf{R}}_{m,k}[f]\right) 
\end{align}
Assuming uniform priors over all states, the posterior probability
of state $s$ is given by $\gamma_{s}[n,f]=p_{s}[n,f]/\left(\sum_{k=1}^{K}p_{k}[n,f]\right).$
Finally, the Bayesian estimate of each source power sequence is given by
\begin{eqnarray}
\hat{\sigma}_{k}^{2}[n,f] & = & \sum_{s=1}^{K}\gamma_{s}[n,f]\sigma_{k|s}^{2}[f].
\end{eqnarray}

\section{Speech Separation Experiments}
\label{sec:experiments}
We demonstrate the performance of the proposed method in two scenarios
using two different parameter estimation methods. We report the results
using the signal-to-distortion ratio (SDR) criterion \cite{vincent2006performance}:
\begin{equation}
\mathrm{SDR}_{m,k}=10\log_{10}\frac{\sum_{\tau}\left|\mathbf{c}_{\mathrm{d},m,k}[\tau]\right|^{2}}{\sum_{\tau}\left|\hat{\mathbf{c}}_{\mathrm{d},m,k}[\tau]-\mathbf{c}_{\mathrm{d},m,k}[\tau]\right|^{2}}.
\end{equation}

\subsection{SiSEC ASY}

To understand the performance of the proposed resampling-free source
separation method, we first compare it to resampling-based methods.
In this section, we describe our contribution to the 2018 Signal Separation
Evaluation Campaign (SiSEC) asynchronous source separation (ASY) task
\cite{liutkus2018sisec}. We show results for Task 2, which is to
separate either $K=3$ or $K=4$ talkers from recordings made by $M=4$
portable recorders with two microphones each. 

\begin{table}
\begin{centering}
\begin{tabular}{cccrrcrr}
\toprule
& &  & \multicolumn{2}{c}{Resampled} &  & \multicolumn{2}{c}{Not Resampled}\\
Type & Mics &  & $K=3$ & $K=4$ &  & $K=3$ & $K=4$\\
\midrule
\multicolumn{2}{c}{Unprocessed} &  & $-3.0$ & $-5.0$ &  & $-3.0$ & $-5.0$\\
\midrule
\multirow{2}{*}{Static} & 2 &  & 0.7 & 0.3 &  & 0.7 & 0.3\\
& 8 &  & \textbf{8.2} & \textbf{2.9} &  & 2.1 & 0.1\\
\midrule
\multirow{2}{*}{Varying} & 2 &  & 1.3 & 0.5 &  & 1.3 & 0.5\\
& $2\times4$ &  & 5.5 & 2.2 &  & \textbf{5.5} & \textbf{2.2}\\
\bottomrule
\end{tabular}
\par\end{centering}
\caption{\label{tab:sisec}Mean SDR performance, in dB, of several filters
on the SiSEC ASY dev2 dataset \cite{liutkus2018sisec}.}
\end{table}

Because the sources and microphones are fixed in this scenario, it
is possible to estimate the sample rate offsets and correct for them
before applying ordinary synchronous blind source separation techniques.
Two of the three contributions to SiSEC 2015 ASY, which used the same
data set, adopted this approach \cite{ono20152015}. Our baseline
resampling implementation combines these two approaches from SiSEC
2015: first, we use two-stage correlation maximization \cite{wang2016correlation}
to estimate the sample rate offsets, then correct them using Lagrange
interpolation \cite{markovich2012blind}. The sources are blindly
separated using offline independent vector analysis \cite{ono2011stable},
and we infer the sources' rank-one covariance matrices from the resulting
unmixing filters. We use these blindly estimated covariance matrices
to design the four separation filters compared in the rows of Table \ref{tab:sisec}:
separate static two-channel Wiener filters for each recorder; a
single static Wiener filter using all eight microphones; separate time-varying two-channel filters for each recorder; and finally the
proposed method, with four time-varying two-channel filters designed
using a common set of estimated source power sequences. Each filter is tested with and without resampling the signals.

When the signals are resampled before separation, the synchronous
eight-channel filter outperforms all other methods. When we restrict
the filters to use two microphones, the separation problem is underdetermined,
so the time-varying filters perform better than the static filter.
In fact, when using the other recorders to classify the active
source, the two-channel filter performs nearly as well as the static eight-channel
filter. Next, we test the four filters without resampling the
signals. The two-channel filters are not affected, since the two microphones
of each recorder are synchronously sampled. The eight-channel filter
performs much worse since it relies on across-array coherence. The proposed asynchronous time-varying filter performance
is identical with or without resampling, suggesting that it is resilient
to sample rate offsets. 

\subsection{Cocktail party scenario with moving
wearable arrays}
\label{sec:Moving-Wearable-Arrays}

The proposed method performs worse than previously proposed methods
for the SiSEC ASY data set, which is amenable to resampling, but it should
be better suited to moving arrays for which resampling is difficult
or impossible. We now consider a listening enhancement experiment
in which microphone arrays are attached to moving human listeners in a cocktail
party scenario. In this scenario there are up to eight simultaneous
speech sources. Since current blind source separation techniques are
poorly suited to such large mixtures, and since we wish to demonstrate
the achievable performance of an asynchronous array system, we use
measured rather than estimated spatial parameters for this experiment.

\begin{figure}
\begin{centering}
\hfill \includegraphics{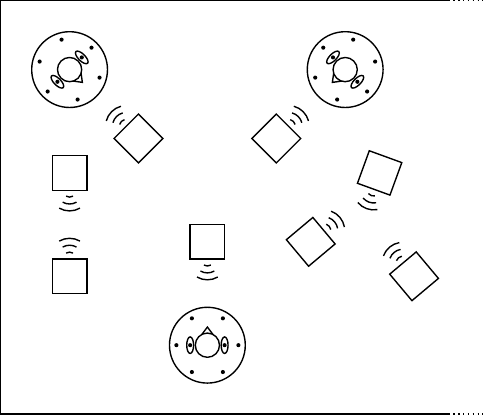} \hfill \includegraphics[height=4.2cm]{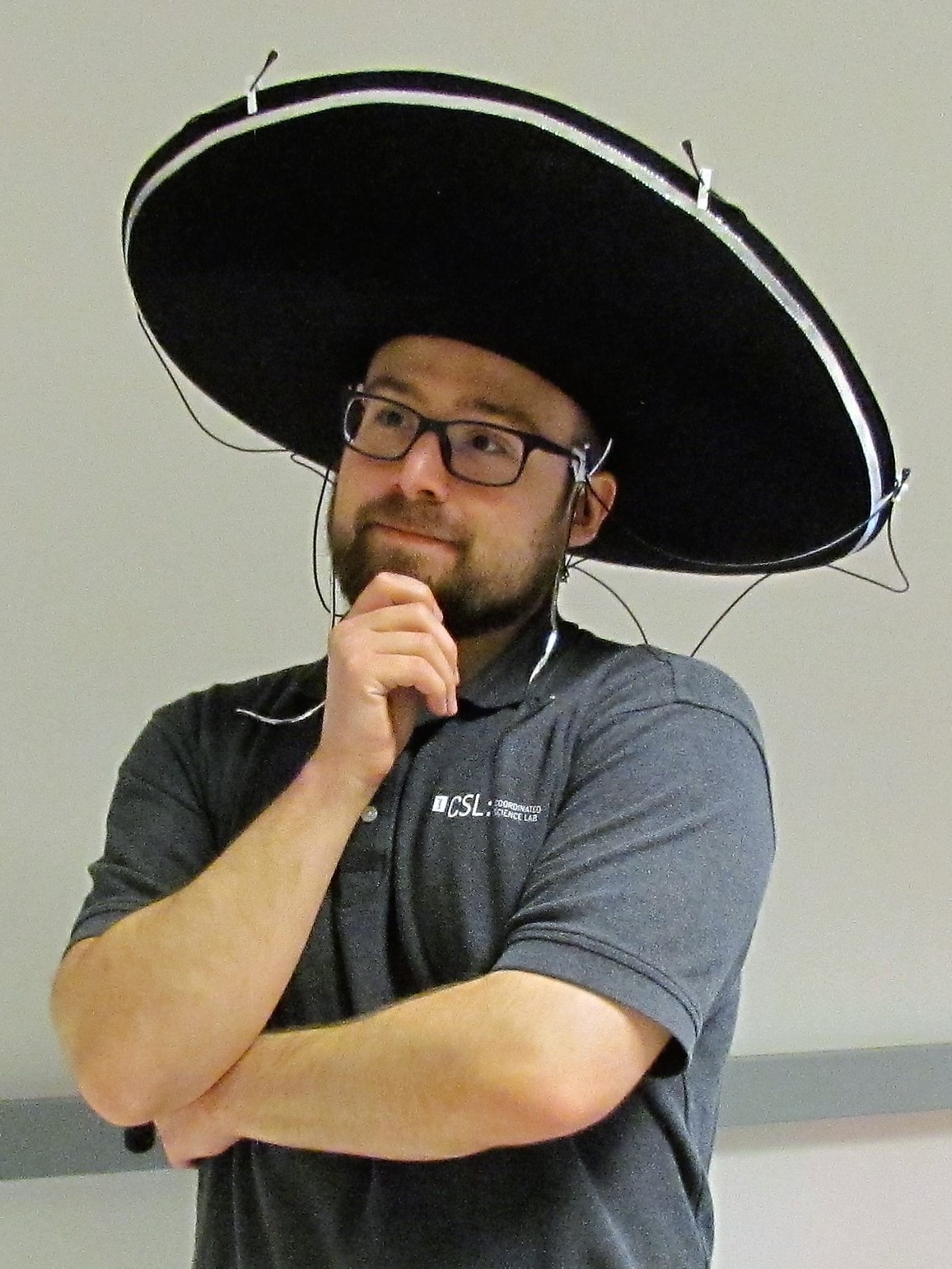} \hfill 
\par\end{centering}
\caption{\label{fig:Cocktail-party-experiment}Left: Cocktail party layout with eight loudspeakers
and three human listeners. Right: Each human
listener wears two in-ear microphones and a hat with six additional microphones.}
\end{figure}

The recordings were made in the Augmented Listening Laboratory at
the University of Illinois at Urbana-Champaign, which has a reverberation
time of about $T_{60}=300$ ms. The cocktail party scenario, shown
in Figure \ref{fig:Cocktail-party-experiment}, consists of eight
talkers, which were simulated using loudspeakers playing clips from
the VCTK anechoic speech database \cite{veaux2017cstr}, and three
real human listeners. Each human listener wore a head-mounted array
of eight omnidirectional lavalier microphones: one in each ear and
six affixed to a rigid, wide-brimmed hat with diameter 60 cm. The
listeners moved their heads continuously during the recordings, alternately nodding, 
looking around the room, and shifting from side to side.

The twenty-four signals were recorded on a single interface, sampled at
at 16 kHz, and highpass filtered from 100 Hz to remove low-frequency
ambient noise. Artificial sample rate offsets of $\pm0.3$ Hz were
applied to two arrays using Lagrange interpolation \cite{markovich2012blind}.
The STFT was computed with a length-4096 von Hann window 
and 75\% overlap. The spatial covariance matrices $\bar{\mathbf{R}}_{m,k}[f]$ were
estimated using 5-second training clips from the same talkers and with similar listener motion as the 15-second
test clips. Because they are designed for binaural listening
devices, the filters produce only the source image estimates for the microphones
in the ears, not for those on the hat. To
measure the source images, the source signals were recorded individually and then
superimposed to form a mixture. This procedure allows us to measure
the ground truth SDR, but it is physically unrealistic because the
human motion is different in every source recording.\footnote{Separated sound samples using real simultaneous recordings are available on the first author's website.} 

\begin{figure}
\begin{centering}
\includegraphics{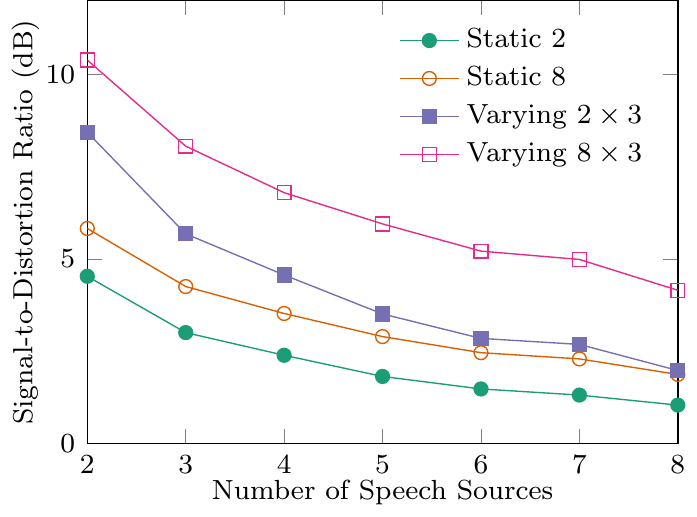}
\par\end{centering}
\caption{\label{fig:cpp_results}Experimental results for a cocktail party
scenario with moving wearable microphone arrays. The SDR is averaged over the left and right ears of all three listeners and over all sources.}
\end{figure}

Figure \ref{fig:cpp_results} compares the separation performance
of four arrays: a static array of two in-ear microphones, a static
array of all eight microphones, a time-varying asynchronous array
of two microphones per listener, and a time-varying asynchronous array
of eight microphones per listener. It is noteworthy that the distributed
array of two microphones per listener outperforms the eight-microphone
static array, even when there are eight sources. The distributed classifier
helps to resolve ambiguities between sources that have similar transfer
functions to the individual arrays. It is particularly important for moving
arrays: when a listener turns their head from side to side, the classifier
can use the other two arrays to decide which source they are hearing.
This feature requires no explicit modeling of head motion; it is a
consequence of the full-rank spatial covariance model and conditional
independence between subarrays.

\section{Conclusions}

The experimental results from Section \ref{sec:experiments} show
that the proposed asynchronous separation method can effectively separate
speech mixtures even when there are more sources than microphones
on each device. The SiSEC results show that it does not perform as
well as a synchronized stationary array, but it does outperform a
single device and does not require sample rate offset estimation or
resampling. The results from the cocktail party scenario show that
the time-varying filters and state classifier work with moving microphones
and scale well to larger arrays. The distributed classifier is
particularly useful for resolving ambiguities when the arrays move
or when sources are far away.

The time-varying filters and classifier both rely on accurate estimation
of the source spatial covariances. In this work, we have not proposed
a method to estimate these parameters without either resampling-based
blind source separation or training data, nor do we explicitly model
their variation over time; asynchronous parameter estimation and tracking
remain important challenges for future work. The proposed asynchronous source
separation system is well suited to distributed arrays in which individual
devices have multiple microphones, are far apart, and are mobile.

\bibliographystyle{ieeetr}
\bibliography{references}

\end{document}